\documentclass{article}
\usepackage{authblk}
\usepackage[utf8]{inputenc}
\usepackage{amsmath}
\usepackage[english]{babel}
\usepackage{titling}
\usepackage{float}
\usepackage[letterpaper,top=2cm,bottom=2cm,left=3cm,right=3cm,marginparwidth=1.75cm]{geometry}
\usepackage{mathrsfs}
\usepackage{graphicx}
\usepackage{siunitx}
\usepackage{booktabs}
\usepackage{physics}
\usepackage{float}
\usepackage[colorlinks=true, allcolors=blue]{hyperref}
\usepackage{cite}
\usepackage{soul}
\usepackage{xcolor}
\usepackage{color}
\usepackage{pdfpages}

\usepackage{caption}                               
\begin{document}

\includepdf[pages={1-18},pagecommand={},width=\textwidth]{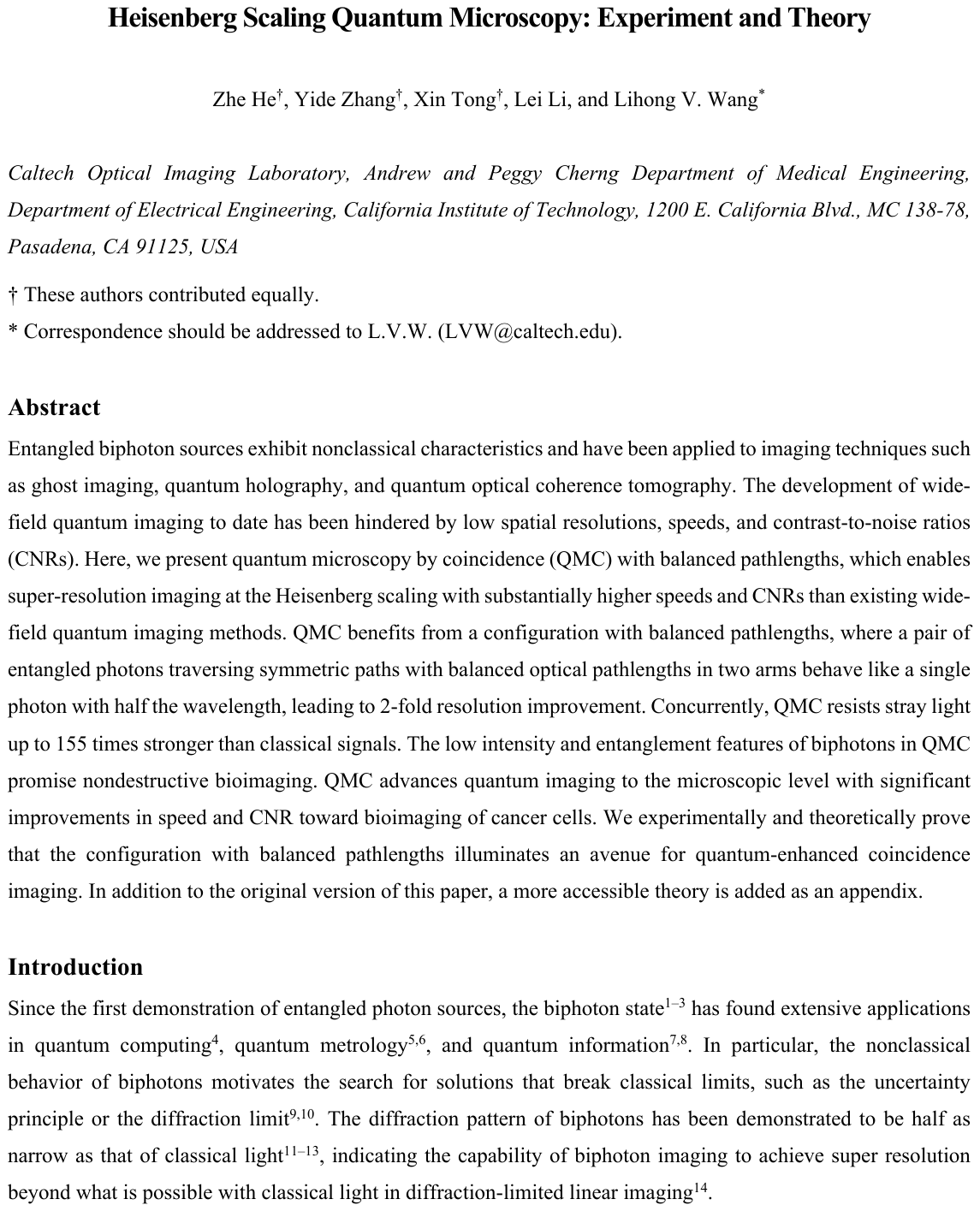}
  
\section*{Supplementary note 1: Theory for super-resolution quantum coincidence imaging}
We compare classical and quantum imaging of a pinhole object through a thin lens, for simplicity, with a magnification of unity. The image of the pinhole yields the point spread function (PSF), defining the spatial resolution. The key difference arises from phase accumulation and quantum interference in the image plane. Because of the approximate transverse shift invariance \cite{goodman2005introduction}, the pinhole is conveniently placed on the optical axis without losing generality.

\subsection*{Classical imaging} 
As shown in Supplementary Fig. \ref{fig:theory1}a, a single photon passes through a thin lens. The classical PSF is the Airy disk determined by the Fourier transform of the circular exit pupil \cite{goodman2005introduction,born2013principles}:
\begin{equation}
I_{\text{CI}}(r) = \left| \int_{\abs{\mathbf{q}} \leq q_{\max}} e^{i \mathbf{q} \cdot \mathbf{r}} d^2\mathbf{q} \right|^2 \propto \left[ \operatorname{somb} \left( q_{\max} r \right) \right]^2,
\end{equation}
where $\mathbf{q}$ denotes the transverse wave vector within the following limit (see Eq.~\ref{eq:qmax}):
\begin{equation}
q_{\max}=k\,\mathrm{NA},\quad k=2\pi/\lambda,\;\mathrm{NA}\le1.
\label{eq:qmaxCI}
\end{equation}
NA denotes the numerical aperture of the circular exit pupil. The classical resolution limit is approximately
\begin{equation}
\Delta r_{\text{CI}} = \frac{ \lambda}{2\, \text{NA}}.
\end{equation}

\subsection*{Quantum imaging with anticorrelation}
In QMC, as shown in Supplementary Fig. \ref{fig:theory1}b, two entangled photons pass through identical lenses and are detected jointly. The following derivation was inspired by Shih \cite{shih2007quantum}.

We begin with a general entangled state of two photons on the object and reference planes in the two arms:
\begin{equation}
\ket{\Psi_{\rm in}}
=
\int_{\lvert\mathbf q\rvert\le q_{\rm src}} d^2\mathbf q\,\Phi(\mathbf q)\,
\hat a_1^\dagger(\mathbf q)\,\hat a_2^\dagger(-\mathbf q)\,\ket{0}
\end{equation}
The transverse-momentum amplitude \(\Phi(\mathbf q)\) is supported on \(\lvert\mathbf q\rvert\le q_{\rm src}\), and \(\hat a_i^\dagger(\mathbf q)\) creates a photon of transverse momentum \(\mathbf q\) in arm \(i=1,2\) while satisfying
\begin{equation}
\bigl[\hat a_i(\mathbf q_i),\hat a_j^\dagger(\mathbf q_j)\bigr]
=\delta_{ij}\,\delta^{(2)}(\mathbf q_i-\mathbf q_j).
\end{equation}
Because the imaging optics in each arm admit only momenta within \(q_\mathrm{pupil}\), we define a projector
\begin{equation}
\hat P
=
\int_{\lvert\mathbf q\rvert\le q_\mathrm{pupil}} d^2\mathbf q\;
\bigl|\mathbf q,\,-\mathbf q\bigr\rangle
\bigl\langle\mathbf q,\,-\mathbf q\bigr|
\quad,\quad
\bigl|\mathbf q_1,\mathbf q_2\bigr\rangle
=\hat a_1^\dagger(\mathbf q_1)\,\hat a_2^\dagger(\mathbf q_2)\,\ket{0}.
\end{equation}

Acting on \(\ket{\Psi_{\rm in}}\), \(\hat P\) truncates support within the following limit (see Eq. \ref{eq:qmaxCI}):
\begin{equation}
    q_\mathrm{max}=\min(q_{\rm src},q_\mathrm{pupil}).
    \label{eq:qmax}
\end{equation}
Consequently, we obtain
\begin{equation}
\ket{\Psi_{\rm img}}
=\hat P\,\ket{\Psi_{\rm in}}
=
\int_{\lvert\mathbf q\rvert\le q_\mathrm{max}} d^2\mathbf q\,\Phi(\mathbf q)\,\hat a_1^\dagger(\mathbf q)\,\hat a_2^\dagger(-\mathbf q)\,\ket{0}.
\end{equation}

On the image plane, the positive-frequency field at transverse position \(\mathbf r\) is
\begin{equation}
\hat E_i^{(+)}(\mathbf r)
=\int d^2\mathbf q\,\hat a_i(\mathbf q)\,e^{i \mathbf q\cdot\mathbf r}.
\end{equation}
The joint two-photon detection operator is
\(\hat\Psi(\mathbf r_1,\mathbf r_2) = \hat E_1^{(+)}(\mathbf r_1)\,\hat E_2^{(+)}(\mathbf r_2)\).

The two-photon wavefunction gives the probability amplitude to detect one photon at \(\mathbf r_1\) in arm 1 and one at \(\mathbf r_2\) in arm 2:
\begin{equation}
\psi(\mathbf r_1,\mathbf r_2)
=
\bra{0}\,\hat\Psi(\mathbf r_1,\mathbf r_2)\,\ket{\Psi_{\rm img}}
=
\bra{0}\,\hat E_1^{(+)}(\mathbf r_1)\,\hat E_2^{(+)}(\mathbf r_2)\,
\int_{\lvert\mathbf q\rvert\le q_{\max}} d^2\mathbf q\,\Phi(\mathbf q)\,\hat a_1^\dagger(\mathbf q)\,\hat a_2^\dagger(-\mathbf q)\,\ket{0}.
\end{equation}
Substituting the field operators yields

\begin{equation}
\begin{aligned}
\psi(\mathbf r_1,\mathbf r_2)
&=
\int d^2\mathbf q_1\,d^2\mathbf q_2\,e^{i(\mathbf q_1\cdot\mathbf r_1+ \mathbf q_2\cdot\mathbf r_2)}
\int_{\lvert\mathbf q\rvert\le q_{\max}} d^2\mathbf q\,\Phi(\mathbf q)
\\
&\quad\times
\bra{0}\,
\hat a_1(\mathbf q_1)\,\hat a_2(\mathbf q_2)\,
\hat a_1^\dagger(\mathbf q)\,\hat a_2^\dagger(-\mathbf q)\,
\ket{0}.
\end{aligned}
\end{equation}

Using
\begin{equation}
\hat a_1(\mathbf q_1)\,\hat a_1^\dagger(\mathbf q) = \delta^{(2)}(\mathbf q_1-\mathbf q)
\end{equation}
and
\begin{equation}
\hat a_2(\mathbf q_2)\,\hat a_2^\dagger(-\mathbf q) = \delta^{(2)}(\mathbf q_2+\mathbf q),
\end{equation}
we reach

\begin{equation}
\psi(\mathbf r_1,\mathbf r_2)
=
\int_{\lvert\mathbf q\rvert\le q_{\max}} d^2\mathbf q\,\Phi(\mathbf q)\,
e^{i\,\mathbf q\cdot(\mathbf r_1- \mathbf r_2)}.
\end{equation}

If \(\Phi(\mathbf q)\equiv 1\) over the disk, we obtain

\begin{equation}
\psi(\mathbf r_1,\mathbf r_2) = \int_{\lvert\mathbf q\rvert\le q_{\max}} d^2\mathbf q\,e^{i\mathbf q\cdot(\mathbf r_1- \mathbf r_2)}.
\end{equation}

For center-symmetric coincidence detection, we set \(\mathbf{r}_{1} = \mathbf{r}\), \(\mathbf{r}_{2} = -\mathbf{r}\). Then
\begin{equation}
\psi(\mathbf{r}, -\mathbf{r}) = \int_{\lvert\mathbf q\rvert\le q_{\max}} d^2\mathbf q\,e^{2i \mathbf q\cdot\mathbf r}.
\end{equation}

The coincidence detection is described by the second-order correlation function:
\begin{equation}
G_{\mathrm{QMC}}^{(2)}(\mathbf{r}, -\mathbf{r}) = |\psi(\mathbf{r}, -\mathbf{r})|^2 
\propto I_{\text{CI}}(2r)
\end{equation}
Hence, the quantum PSF is an Airy pattern that is narrowed by a factor of 2.
\begin{equation}
\Delta r_{\text{QMC}} = \frac{1}{2} \Delta r_{\text{CI}} = \frac{ \lambda}{4\, \text{NA}}.
\end{equation}
This confirms that the coincidence imaging achieves super-resolution consistent with the experimental observations, arising from transverse phase accumulation in the two-photon wavefunction.

\begin{figure}[H]
    \centering
    \includegraphics[width=\textwidth]{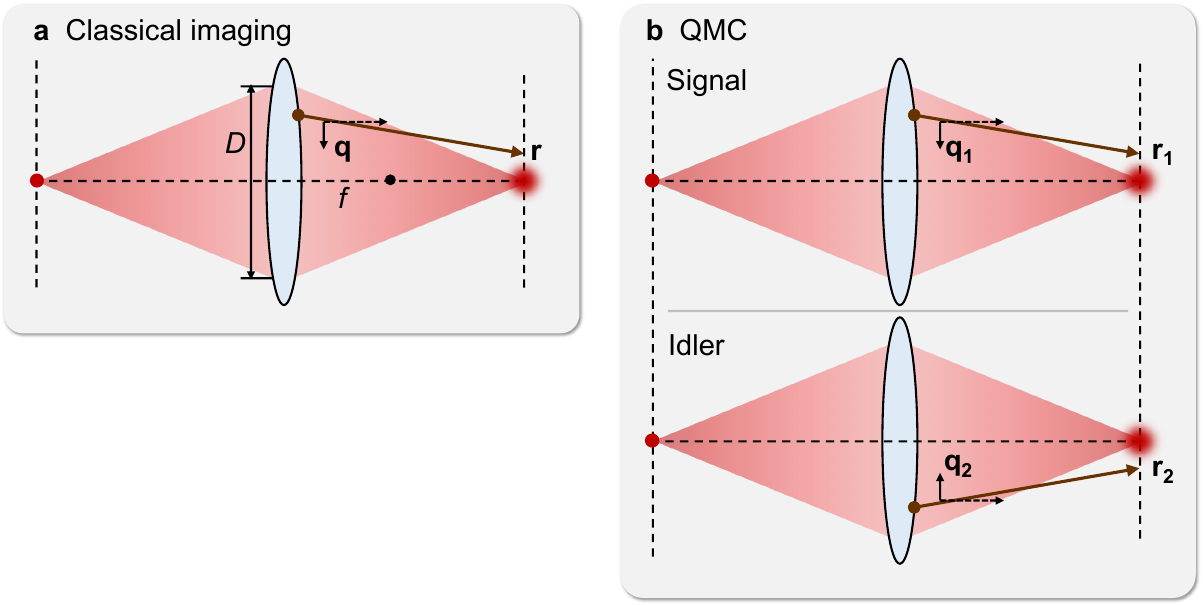}
    \caption{Simplified schematics of (\textbf{a}) classical imaging and (\textbf{b}) QMC.}
    \label{fig:theory1}
\end{figure}

\subsection*{Quantum imaging with correlation}
The above derivation applies to our far-field experimental implementation, where the momentum space of the source plane is mapped onto both the object and reference planes. Alternatively, the source plane can be conjugated directly to the object and reference planes, a configuration sometimes referred to as near-field imaging.

In this case, the image-plane wavefunction becomes
\begin{equation}
\psi(\mathbf{r}_1, \mathbf{r}_2) 
= \iint_{|\mathbf{q}| \leq q_{\max}} d^2\mathbf{q}_1 \, d^2\mathbf{q}_2 \, \delta(\mathbf{q}_1 - \mathbf{q}_2) \, e^{i \mathbf{q}_1 \cdot \mathbf{r}_1} e^{i \mathbf{q}_2 \cdot \mathbf{r}_2} 
= \int_{|\mathbf{q}| \leq q_{\max}} d^2\mathbf{q} \, e^{i \mathbf{q} \cdot (\mathbf{r}_1 + \mathbf{r}_2)}.
\end{equation}
For photon-number-resolving coincidence detection (\(\mathbf{r}_1 = \mathbf{r}\), \(\mathbf{r}_2 = \mathbf{r}\)), the point spread function given by the second-order correlation function is narrowed by $2\times$ as well:
\begin{equation}
G^{(2)}(\mathbf{r}, \mathbf{r}) 
= \left| \psi(\mathbf{r}, \mathbf{r}) \right|^2 
= \left| \int_{|\mathbf{q}| \leq q_{\max}} d^2\mathbf{q} \, e^{2i \mathbf{q} \cdot \mathbf{r}} \right|^2 
\propto I_{\text{CI}}(2r).
\end{equation}

\subsection*{Quantum imaging with decorrelation} 
Quantum imaging with disrupted momentum correlation becomes equivalent to classical confocal microscopy. The detection-plane two-photon wavefunction becomes
\begin{equation}
\psi(\mathbf{r}_p; \mathbf{r}_1, \mathbf{r}_2) = \iint_{|\mathbf{q}| \leq q_{\max}} d^2\mathbf{q}_1 \, d^2\mathbf{q}_2 \, 
e^{i \mathbf{q}_1 \cdot (\mathbf{r}_1-\mathbf{r}_p)} e^{i \mathbf{q}_2 \cdot (\mathbf{r}_2-\mathbf{r}_p)}, 
\end{equation}
where $\mathbf{r}_p$ denotes the pinhole position. For co-axial coincidence detection ($\mathbf r_{1}=\mathbf 0, \mathbf r_{2}=\mathbf 0$),
\begin{equation}
\psi(\mathbf{r}_p; \mathbf 0,\mathbf 0)
=\Bigl(\int_{|\mathbf{q}| \leq q_{\max}} d^{2}\mathbf q \, e^{-i\mathbf {q} \cdot \mathbf{r}_p}\Bigr)
\;\times\;
\Bigl(\int_{|\mathbf{q}| \leq q_{\max}} d^{2}\mathbf q \, e^{-i\mathbf q \cdot \mathbf{r}_p}\Bigr).
\end{equation}
The PSF is narrowed by approximately $\sqrt{2}\times$:
\begin{equation}
G^{(2)}(\mathbf{r}_p; \mathbf 0,\mathbf 0)
=\lvert\psi(\mathbf{r}_p; \mathbf 0,\mathbf 0)\rvert^{2}
\propto \bigl[I_{\mathrm{CI}}(r_p)\bigr]^{2}.
\end{equation}

\includepdf[pages={1-8},pagecommand={},width=\textwidth]{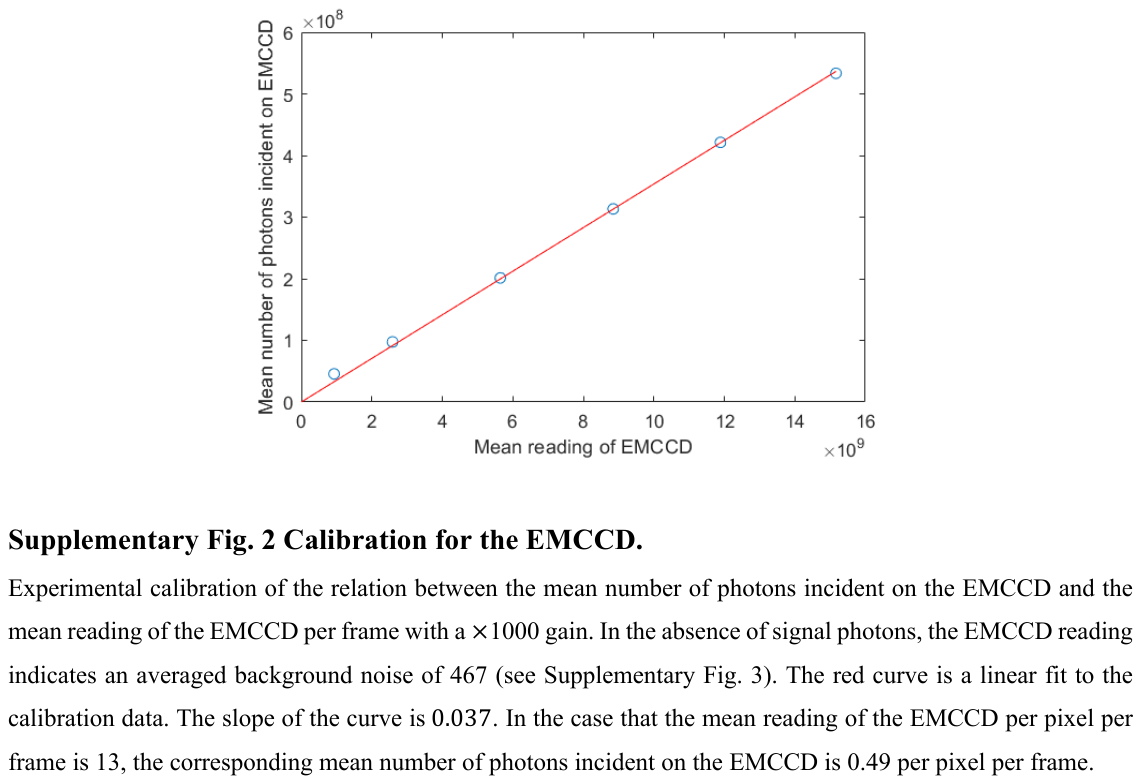}

\bibliographystyle{unsrt}
\bibliography{references}
\end{document}